\documentclass[prd,aps,preprint,tightenlines,superscriptaddress]{revtex4}
\usepackage{epsfig}
\usepackage{amsmath}
\begin{document}

\preprint{RBRC-495} \preprint{DOE/40762-345}

\title{Transverse Momentum Distribution Through \\
Soft-Gluon Resummation in Effective Field Theory}

\author{Ahmad Idilbi}
\email{idilbi@physics.umd.edu} \affiliation{Department of Physics,
University of Maryland, College Park, Maryland 20742, USA}
\author{Xiangdong Ji}
\email{xji@physics.umd.edu} \affiliation{Department of Physics,
University of Maryland, College Park, Maryland 20742, USA}
\author{Feng Yuan}
\email{fyuan@quark.phy.bnl.gov} \affiliation{RIKEN/BNL Research
Center, Building 510A, Brookhaven National Laboratory, Upton, NY
11973}

\date{\today}
\vspace{0.5in}
\begin{abstract}
We study resummation of transverse-momentum-related large
logarithms generated from soft-gluon radiations in soft-collinear
effective field theory. The anomalous dimensions of the effective
quark and gluon currents, an important ingredient for the
resummation, are calculated to two-loop order. The result at
next-to-leading-log reproduces that obtained using the standard
method for deep-inelastic scattering, Drell-Yan process, and Higgs
production through gluon-gluon fusion. We comment on the extension
of the calculation to next-to-next-to-leading logarithms.
\end{abstract}

\maketitle

\section{Introduction}

Perturbative calculations of high-energy processes with widely
separated scales often yield large logarithms involving the ratios
of the scales, which shall be resummed to all orders to achieve
reliable predictions. In processes like deep-inelastic scattering
(DIS), Drell-Yan (DY) and the Higgs-boson production, multiple
hard scales appear in a certain kinematical limit such as
threshold production region \cite{Ste87,{CatTre89}}, and/or when a
low transverse momentum \cite{DokDyaTro80,{ParPet79}} of the final
states is measured. For the transverse momentum distribution, the
rigorous theoretical study in QCD started with the classical work
on semi-inclusive processes in $e^+e^-$ annihilation by Collins
and Soper in \cite{ColSop81}, where a factorization was proved
based on the transverse momentum-dependent (TMD) parton
distributions and fragmentation functions \cite{ColSop81p}. The
resummation of TMD large logarithms was performed by solving the
relevant energy evolution equation. This approach was later
applied to the DY process in \cite{ColSopSte85}, where a general
and systematic analysis of the factorization and resummation were
performed. This latter procedure became known as
``Collins-Soper-Sterman'' (CSS) resummation formalism. The
resummed formulas are used for many processes with the relevant
coefficients extracted by comparing between the expansion of the
resummed expressions and the fixed-order calculations
\cite{davies}. Although the factorization approach is sound and
rigorous, it involves often quantities that are not manifestly
gauge invariant. Moreover, the physics of the energy evolution
does not seem transparent.

In this paper we pursue the same resummation from a different
path. We exploit the fact that there are (at least) two well-
separated hard scales which are naturally appropriate for an
effective field-theoretic approach to perform the resummation.
The recently proposed ``soft-collinear-effective-theory'' (SCET)
\cite{SCET} is useful here. Although it was originally applied for
the study of heavy $B$ meson decays, it was later generalized to
other high-energy processes \cite{SCET1}. Recently, this effective
theory has been used to study the threshold resummation for the
DIS structure function as $x\to 1$ in \cite{Man03} and for DY
\cite{IdiJi05}. The resummation in the effective theory is
performed by studying the anomalous dimensions $\gamma_1$ of the
effective operators after performing matching between the full and
effective theories. The exponentiated Sudakov form factor appears
by running down the scale of the matching coefficient from the
higher scale $\mu_H^2\sim Q^2$ down to the lower scale $\mu_L^2
\sim \lambda Q^2$ ($\lambda\ll 1$).

A SCET study of transverse-momentum dependence for Higgs-boson
production was initiated in Refs. \cite{Lics05}, where the authors
work directly in momentum space, and the next-to-leading
logarithmic result was deduced from a full QCD calculation. Here
we consider the same resummation for the Drell-Yan process as well
as the standard-model Higgs production at hadron colliders.
Instead of working in the momentum space, we use the impact
parameter space. Moreover we perform calculations directly in
SCET.

In general, the resummation formula for these processes can be
written in the following form, taking the DY process as an example
\cite{ColSopSte85},
\begin{equation}
\frac{d\sigma}{dQ^2dydQ_T^2}=\sigma_0\left[\int\frac{d^2\vec{b}}
{(2\pi)^2}e^{-i\vec{Q}_T\cdot \vec{b}}~
W(b,Q^2,x_1,x_2)+Y(Q_T,Q^2, x_1,x_2)\right] \ ,
\end{equation}
where $\sigma_0=4\pi^2\alpha_{\rm em}^2/(9sQ^2)$ represents the
Born-level cross section for DY production. The variables used
here are standard: $Q^2$ is the invariant mass of the DY pair;
$Q_T$ is the observed transverse momentum relative to the beam
axis; $y=(1/2)\ln (Q^0+Q^3)/(Q^0-Q^3)$ is the rapidity;
$s=(P_1+P_2)^2$ is the center-of-mass energy squared with $P_i$
the momentum of the incoming hadrons. The variable $x_1$ and $x_2$
are the equivalent parton fractions, $x_1 = \sqrt{Q^2/s} ~e^y$ and
$x_2 = \sqrt{Q^2/s}~ e^{-y}$. The $W$ term contains the most
singular contributions at small $Q_T$, resummed to all orders in
perturbation theory. The second term $Y(Q_T,Q^2)$ represents the
regular part of a fixed-order calculation for the cross section,
which becomes important when the transverse momentum $Q_T$ is on
the order of $\sqrt {Q^2}$.

The main result of our study can be summarized by the following
formula for the $W$ term,
\begin{eqnarray}
W(b,Q^2,x_1,x_2)&=&\sum_{qq'}C^2(Q^2,\alpha_s(Q^2))e^{-{\cal
S}(Q^2,~\mu_L^2)} \nonumber \\
&& \times \left( C_q\otimes q\right) (x_1,b,Q,\mu_L^2)\times
\left( C_{q'}\otimes {q'}\right) (x_2,b,Q,\mu_L^2) \ ,
\end{eqnarray}
where $q$ and $q'$ are the parton distributions and/or
fragmentation functions related to the processes studied, and the
notation $\otimes$ stands for convolution. Two matching
coefficients appear in the above formula: one is
$C(Q^2,\alpha_s(Q^2))$ connecting between the matrix element of
full QCD current and the effective theory analogue at the scale
$Q^2$; the other is the coefficient function $C_q$ obtained by
calculating the processes in SCET at a lower scale $\mu_L^2 \sim
Q_T^2$. The exponential suppression form factor arises from the
anomalous dimension of the effective current in SCET,
\begin{equation}
{\cal S}(Q^2,\mu_L^2)=\int_{\mu_L}^Q \frac{d\mu}{\mu}
2\gamma_1(\mu^2,\alpha_s(\mu^2)) \
\end{equation}
The anomalous dimension is identical for DIS and DY processes, and
depends only on the effective theory operators. Moreover, the same
$\gamma_1$ controls the threshold and the low transverse momentum
resummations. The matching coefficient $C(Q^2,\alpha_s(Q^2))$, on
the other hand, is process dependent (but independent of threshold
or transverse-momentum resummation).
All the large double logarithms are included in the Sudakov form
factor ${\cal S}$.

In the main body of the paper, we show how to get the above
resummation formula using SCET.
In Sec.~II, we consider the resummation for the DY process in
SCET. The anomalous dimensions of the effective current are
calculated up to two-loop order. A comparison will be made with
the CSS formalism. In Sec. III we will briefly discuss the
extension of the formalism to the standard-model Higgs production.
We conclude the paper in Sec. IV.

\section{Drell-Yan production at low $Q_T$ in SCET}

In soft-collinear effective field theory, we consider Drell-Yan
production with finite transverse momentum $Q_T$ in two steps,
assuming $Q\gg Q_T\gg\Lambda_{\rm QCD}$. In the first step, one
integrates out all loops with virtuality of order $Q$ to get an
effective theory called SCET$_{\rm I}$ in which there are only
collinear and soft modes. The collinear modes have virtuality of
order $Q_T$. In the second step, one integrates out the collinear
modes with virtuality $Q_T$. In this case, the theory is matched
onto SCET$_{\rm II}$ which is just the ordinary QCD without
external hard scales. The soft physics is now included in the
parton distribution functions.

Let us consider the first step: integrating out modes of
virtuality of order $Q$. At low $Q_T$, the most singular
contribution in DY comes from the form-factor type of diagrams, in
which the quark and antiquark first radiate soft gluons, followed
by an annihilation vertex decorated with loop corrections. If the
gluon radiations are attached to the loops, the soft gluon limit
does not give rise to any infrared singularity, and hence diagrams
yield higher-order contributions in $Q_T/Q$. Therefore, one needs
to consider only the form-factor type of diagrams in studying the
virtual corrections. To integrate out the hard modes (where all
gluon momenta are of order $Q$-- see, e.g., \cite{Zheng} ) from
the theory, we can match the full QCD current onto the (gauge
invariant) SCET current \cite{Man03,IdiJi05} and the matching
coefficient contains the hard contributions. By exploiting the
non-renormalizability of the form factor, we write down a simple
renormalization group equation from which we can extract the
anomalous dimension of the effective current.

In the second step, we calculate the SCET cross section. One needs
to compute only the real contributions since the virtual diagrams
are scaleless in the effective theory and vanish in on-shell
dimensional regularization (DR). At this stage the cross section
obtained is the same as the one in full QCD taken to the relevant
kinematical limit. This has been established at one-loop order in
the threshold resummation for DIS and DY \cite{Man03,IdiJi05}.
This will also be the case for DY in low transverse momentum
limit.


Resumming the large logarithmic ratios is performed by considering
the scale dependence of the matching coefficient, which is
controlled by the anomalous dimension of the effective current. By
running down the scale from $\mu_H^2\sim Q^2$ to the lower scale
$\mu_L^2$, all the large logarithms exponentiate and give rise to
the Sudakov form factor. In the following, we will demonstrate how
this can be done systematically, providing a powerful tool for
future studies. The basic lagrangian and Feynman rules for SCET
can be found in Refs.\cite{SCET}. In our calculation, we choose
the two light-like vectors: $n=1/\sqrt{2}(1,0,0,-1)$ and $\bar
n=1/\sqrt{2}(1,0,0,1)$. The matching is made between the full QCD
current $\overline{\psi}\gamma_\mu\psi$ and the SCET$_{\rm I}$
current $\bar \xi_{n}W_{n}\gamma_\mu W_{\bar n}^\dagger \xi_{\bar
n}$, where $\xi_{n,\bar n}$ are the collinear quark fields in
SCET, and $W_{n,\bar n}$ are the collinear Wilson lines. The
collinear Wilson lines appear as a requirement of the collinear
gauge invariance in the SCET$_{\rm I}$ lagrangian. All
calculations are performed in Feynman gauge and ${\overline{\rm
MS}}$ scheme in $d=4-2\epsilon$. DR is employed to regularize both
infrared (IR) as well as ultraviolet (UV) divergences.

\subsection{Matching at scale $Q^2$ and anomalous
dimension of SCET current at two loops}

At the scale of $Q^2$, we need to consider only the virtual
contributions to the quark form factor both in full QCD and the
effective theory. The full QCD calculation of the quark form
factor has so far been done up to two-loop order \cite{van}. On
the other hand, the virtual diagrams in the effective theory are
scaleless, and the relevant form factors vanish in DR where the UV
and IR divergences cancel out at every order in $\alpha_s$. Since
the effective theory captures the IR behavior of the full theory
(and the UV divergences are cancelled in both theories by
respective counterterms), the matching coefficients for the
effective theory operators will be obtained from the finite parts
of the full QCD calculation. In the following, we will study the
matching coefficients at one- and two-loop orders.

At one-loop order, the matching coefficient has been calculated
($C = \sum_n C^{(n)}$) \cite{Man03,IdiJi05},
\begin{equation}
C^{(1)}(\mu^2,\alpha_s(\mu^2)) =
-\frac{\alpha_s(\mu^2)C_F}{4\pi}\left[\ln^2\frac{Q^2}{\mu^2}-3\ln\frac{Q^2}{\mu^2}
+8-\frac{7\pi^2}{6}\right] \ ,
\end{equation}
where, throughout the paper, $\alpha_s$ is the renormalized
running coupling constant.

From the above result, we can get the order-$\alpha_s$ anomalous
dimension for the SCET current ($\gamma_1 = \sum_n
\gamma_1^{(n)}$),
\begin{equation}
\gamma_1^{(1)}=\frac{\alpha_s(\mu^2)
C_F}{\pi}\left[\ln\frac{Q^2}{\mu^2}-\frac{3}{2}\right] \
\end{equation}
The matching coefficient satisfies the renormalization group
equation,
\begin{equation}
\frac{\rm d}{\rm d\ln\mu}C(\mu, \alpha_s(\mu))=\gamma_1(\mu)C(\mu,
\alpha_s(\mu)) \
\end{equation}
At the scale $\mu_H^2=Q^2$ we have,
\begin{equation}
C(Q^2,\alpha_s(Q^2))=1+C^{(1)}=1-\frac{\alpha_s(Q^2)}{2\pi}\left(4-\frac{7\pi^2}{12}\right
) \
\end{equation}
We note that the matching scale can also be chosen as
$\mu_H^2=C_2^2Q^2$ with $C_2$ a constant of order unity. The $C_2$
dependence of the matching coefficient corresponds to the $C_2$
dependence in the CSS resummation \cite{ColSopSte85}. However, in
order to minimize the logarithms in the matching coefficient
$C(C_2^2Q^2,\alpha_s(C_2^2Q^2))$, the best choice seems to be
$C_2=1$.

To get the anomalous dimension at two-loop order, we need to
calculate the matching coefficient up to the logarithmic term at
the same order. From the quark form factor at two-loop order, we
get the matching coefficient for DY process,
\begin{eqnarray}
C^{(2)}(
Q^2,\mu^2)&=&\frac{\alpha_s^2}{(4\pi)^2}\left\{C_F^2\left[\frac{1}{2}
\left(\ln^2\frac{-Q^2}{\mu^2}-3\ln\frac{-Q^2}{\mu^2}+8-\zeta(2)\right)^2\right.\right.\nonumber\\
&&\left.+\left(24\zeta(3)+\frac{3}{2}-12\zeta(2)\right)\ln\frac{-Q^2}{\mu^2}\right]\nonumber\\
&&\left.+N_fC_F\left[-\frac{2}{9}\ln^3\frac{-Q^2}{\mu^2}+\frac{19}{9}
\ln^2\frac{-Q^2}{\mu^2}-\left(\frac{209}{27}-\frac{4}{3}\zeta(2)\right)\ln\frac{-Q^2}{\mu^2}\right]\right.\nonumber\\
&&\left.+C_AC_F\left[\frac{11}{9}\ln^3\frac{-Q^2}{\mu^2}+\frac{1}{2}\left(4\zeta(2)
-\frac{332}{9}\right)\ln^2\frac{-Q^2}{\mu^2}\nonumber\right.\right.\\
&&\left.\left.-\left(\frac{2545}{54}+\frac{22}{3}\zeta(2)-26\zeta(3)\right)\ln\frac{-Q^2}{\mu^2}\right]
\right\}+\cdots \
\end{eqnarray}
Here $\cdots$ stands for the constant terms which do not
contribute to the anomalous dimension. The result contains an
imaginary part because the DY form factor contains final-state
rescattering. From the above result we can calculate the anomalous
dimension at order $\alpha_s^2$. The expansion of Eq. (6) to order
$\alpha_s^2$ gives
\begin{eqnarray}
\frac{\partial}{\partial\ln\mu} C^{(2)}+\frac{\partial
\alpha_s}{\partial\ln\mu}\frac{\partial}{\partial
\alpha_s}C^{(1)}&=&\gamma_1^{(1)}C^{(1)}+\gamma_1^{(2)}C^{(0)}=\gamma_1^{(1)}C^{(1)}+\gamma_1^{(2)}
\ .
\end{eqnarray}
The running of the strong coupling constant $\alpha_s(\mu)$ is
\begin{equation}
\frac{\partial}{\partial\ln\mu}\alpha_s=-2\beta_s \alpha_s^2 \ ,
\end{equation}
where $\beta_s$ has the following expansion,
\begin{equation}
\beta_s=\frac{1}{4\pi}\left[\beta_0+\frac{\alpha_s}{4\pi}\beta_1+\cdots\right]
\ .
\end{equation}
 $\beta_0$ and $\beta_1$ are defined as
\begin{eqnarray}
\beta_0=\frac{11}{3}C_A-\frac{2}{3}N_f,~~~
\beta_1=\frac{34}{3}C_A^2-4N_fT_FC_F-\frac{20}{3}N_fT_FC_A \ ,
\end{eqnarray}
where $N_f$ is the number of light flavors, $C_A=N_C$ with $N_C$
the number of colors, $C_F=(N_C^2-1)/2N_C$, $T_F=1/2$. Using the
above formulas, we get the anomalous dimension at two-loop order,
\begin{eqnarray}
\gamma_1^{(2)}&=&\left(\frac{\alpha_s}{\pi}\right)^2\left\{\left[\left(\frac{67}{36}-\frac{\pi^2}{12}\right)C_A-
\frac{5}{18}N_f\right]C_F\ln\frac{Q^2}{\mu^2}\nonumber\right.\\
&&~~~~+\left(\frac{13}{4}\zeta(3)-\frac{961}{16\times
27}-\frac{11}{48}\pi^2\right)
C_AC_F\nonumber\\
&&~~~~+\left(\frac{\pi^2}{24}+\frac{65}{8\times
27}\right)N_fC_F\nonumber\\
&&~~~~\left.+\left(\frac{\pi^2}{4}-
\frac{3}{16}-3\zeta(3)\right)C_F^2 \right\}  \ .
\end{eqnarray}
Using the notation of \cite{vanHiggs}, we can write
\begin{equation}
\gamma_1^{(2)}=\frac{\alpha_s^2}{(4\pi)^2}\left(\frac{1}{2}\gamma_{Kqq}^{(1)}\ln\frac{Q^2}{\mu^2}-
\gamma_{qq}^{(1)}-2f_{q2}^{(1)}\right) \
\end{equation}
where
\begin{equation}
\gamma_{Kqq}^{(1)}=16C_FK,~~~K=C_A\left(\frac{67}{18}-\zeta(2)\right)-\frac{10}{9}N_fT_F\
\end{equation}
is proportional to the anomalous dimension of a Wilson line cusp,
and
\begin{eqnarray}
\gamma_{qq}^{(1)}&=&
   C_F^2\left(3-24\zeta(2)+48\zeta(3)\right)+C_AC_F\left(\frac{17}{3}+\frac{88}{3}\zeta(2)-24\zeta(3)\right)
        \nonumber \\
  && +N_fT_FC_F\left(-\frac{4}{3}-\frac{32}{3}\zeta(2)\right)
\end{eqnarray}
is proportional to the coefficient of $\delta(1-x)$ in the quark
splitting function, and finally
\begin{equation}
f_{q2}^{(1)}=C_F\left[-2\beta_0\zeta(2)
+C_A\left(\frac{404}{27}-14\zeta(3)\right)+N_fT_FC_F\left(-\frac{112}{27}\right)\right]
\ .
\end{equation}

The above calculation can be repeated for the DIS process and the
same anomalous dimension is obtained.

\subsection{Matching at scale $\mu_L^2$}

To perform matching at the lower scale $\mu_L^2\ll Q^2$ we
calculate the cross section in SCET$_{\rm I}$ and match it to a
product of quark distributions. From the result we can extract the
coefficient functions. The parton distributions in SCET$_{\rm II}$
are the same as those in full QCD \cite{Man03,IdiJi05}. Below the
scale $Q^2$, the hard modes have been integrated out, and have
been taken into account by the matching condition at $Q^2$.
Therefore, the calculation of the cross section at $\mu_L^2$ is
performed with SCET$_{\rm I}$ diagrams, including both virtual and
real contributions. As mentioned earlier the virtual diagrams in
SCET are scaleless and vanish in pure DR. As such one can ignore
them, but the counterterm for the effective current must be taken
into account. This is equivalent to taking the UV-subtracted
contribution from the virtual diagrams. The counterterm has been
calculated in \cite{Man03},
\begin{eqnarray}
{\rm c.t.}&=& \frac{\alpha_s}{2\pi} C_F \left [
-\frac{1}{\epsilon^2} - \frac{3}{2\epsilon} - \frac{1}{\epsilon}
\ln \frac{\mu^2}{Q^2} \right ] \ ,\label{contour}
\end{eqnarray}
where $\mu^2$ will be taken as $\mu_L^2$.
\begin{figure}[t]
\begin{center}
\includegraphics[height=2.0cm]{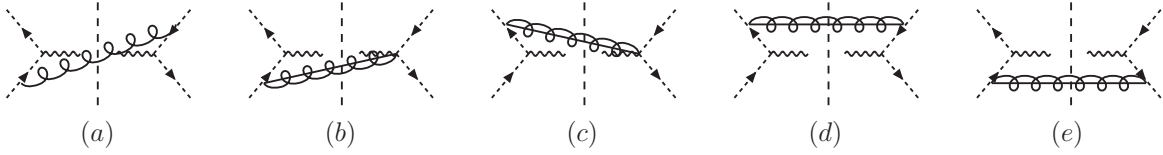}
\end{center}
\vskip -0.7cm \caption{\it Non-vanishing Feynman diagrams
contributing to Drell-Yan production in the
soft-collinear-effective theory: (a) for the soft gluon radiation;
(b)-(e) for $n$ and $\bar n$ collinear gluon radiations. The
mirror diagrams of (a-c) are not shown here but are included in
the results.}
\end{figure}

The real contribution contains collinear and soft gluon radiation
diagrams. Some of these diagrams are identically zero because of
$n^2=0$ or $\bar n^2=0$, or because of the equation of motion for
the external collinear (anti-)quark. The non-vanishing diagrams
are shown in Fig.~1, including soft-gluon interference
contribution, $n$ and $\bar n$ collinear gluon radiations and
their interferences. The contribution from Fig.~1(a) is, including
that from the mirror diagram,
\begin{eqnarray}
\frac{d^3\sigma^{(1)}}{d^2Q_T dy}|_{\rm
fig.1(a)}=\sigma_0\frac{\alpha_sC_F}{\pi^2}\frac{1}{Q_T^2}\left[-\frac{\delta(x_2-1)}{(1-x_1)_+}-
\frac{\delta(x_1-1)}{(1-x_2)_+}-\delta(x_1-1)\delta(x_2-1)\ln\frac{Q^2}{Q_T^2}\right]
\
\end{eqnarray}
Fig.1(b) represents the interference between the $\bar
{n}$-collinear gluon radiation with collinear expansion of the
current operator, and its contribution is,
\begin{eqnarray}
\frac{d^3\sigma^{(1)}}{d^2Q_T dy}|_{\rm
fig.1(b)}=\sigma_0\frac{\alpha_sC_F}{\pi^2}\frac{1}{Q_T^2}\left[\frac{x_1\delta(x_2-1)}{(1-x_1)_+}+
\frac{\delta(x_1-1)}{(1-x_2)_+}+\delta(x_1-1)\delta(x_2-1)\ln\frac{Q^2}{Q_T^2}\right]
\
\end{eqnarray}
while Fig.1(c) for the $n$-collinear gluon radiation yields,
\begin{eqnarray}
\frac{d^3\sigma^{(1)}}{d^2Q_T dy}|_{\rm
fig.1(c)}=\sigma_0\frac{\alpha_sC_F}{\pi^2}\frac{1}{Q_T^2}\left[
\frac{x_2\delta(x_1-1)}{(1-x_2)_+}+\frac{\delta(x_2-1)}{(1-x_1)_+}
+\delta(x_1-1)\delta(x_2-1)\ln\frac{Q^2}{Q_T^2}\right]\
\end{eqnarray}
Figs. 1(d) and (e) stand for the $n$ and $\bar n$ collinear gluon
radiations, respectively, and their sum is,
\begin{eqnarray}
\frac{d^3\sigma^{(1)}}{d^2Q_T dy}|_{\rm
fig.1(de)}=\sigma_0\frac{\alpha_sC_F}{2\pi^2}\frac{1}{Q_T^2}\left(1-\epsilon\right)
\left[(1-x_1)\delta(x_2-1)+(1-x_2)\delta(x_1-1)\right] \ .
\end{eqnarray}
The sum of the above contributions reproduces the result of the
full QCD calculation in the limit of low transverse momentum (see,
e.g., \cite{JiMaYu04}).

From the above, the real contribution contains soft divergences
(i.e, when $Q_{\perp}^2\rightarrow0$), which will be cancelled by
the virtual contribution. To see this cancellation explicitly, we
Fourier-transform the cross section from the transverse momentum
space into the impact parameter $b$-space. The result is $W(b,
Q^2,\mu_L^2)$ in SCET, including real and virtual contributions
(the contourterm, Eq.~(\ref{contour})),
\begin{eqnarray}
&&W(b,Q^2,\mu_L^2, x_1,x_2)\nonumber\\
&&=\delta(x_1-1)\delta(x_2-1) +
\frac{\alpha_sC_F}{2\pi}\left[{\cal P}_{q/q}(x_1)\delta(x_2-1)
+(x_1\leftrightarrow
x_2)\right]\left(-\frac{1}{\epsilon}-\gamma_E+\ln\frac{4}{4\pi \mu_L^2 b^2}\right) \nonumber\\
&&+\frac{\alpha_sC_F}{2\pi}\left[(1-x_1)\delta(x_2-1)+(1-x_2)\delta(x_1-1)\right]
\nonumber\\
&&-\frac{\alpha_sC_F}{2\pi}\delta(x_1-1)\delta(x_2-1)\left[\ln^2\left(\frac{\mu_L^2
b^2}{4}e^{2\gamma_E-\frac{3}{2}}\right)
+2\ln\frac{Q^2}{\mu_L^2}\ln\left(\frac{\mu_L^2b^2}{4}e^{2\gamma_E}\right)-\frac{9}{4}+\frac{\pi^2}{6}\right]
\ , \nonumber\\
\label{21}
\end{eqnarray}
where ${\cal P}_{q/q}$ is the one-loop quark splitting function,
\begin{equation}
{\cal P}_{q/q}(x)=\left(\frac{1+x^2}{1-x}\right)_+ \
\end{equation}
The soft divergences in $W(b,Q^2,\mu_L^2)$ have been cancelled.
There is, however, the collinear divergence left which can be
absorbed into the quark distribution at one-loop order. The cross
section depends on the ultraviolet scale $\mu_L^2$. It is somewhat
surprising that the above result also depends on $\ln Q^2$, but
this is expected from the kinematical constraints of the process.

In order to eliminate the large logarithms in the coefficient
function, the best choice for the scale $\mu_L$ is
$\mu_L^2=C_1^2/b^2$ with $C_1=2e^{-\gamma_E}$. In addition,
$W(b,Q^2,\mu_L^2)$ no longer depends on $Q^2$ at this order. This,
however, may not be true at higher orders because one cannot
eliminate all $\ln Q^2$ by making a single choice of $\mu_L^2$.
Considering the quark distribution at one-loop order, we can cast
Eq. (\ref{21}) into the form,
\begin{equation}
W(b,\mu_L^2=C_1^2/b^2,x_1,x_2)=C_q(b,x_1,\mu_L^2)\otimes
q(x,\mu_L^2)C_{\bar q}(b,x_2,\mu_L^2)\otimes \bar q(x,\mu_L^2) \ ,
\end{equation}
where $C_q$ reads
\begin{eqnarray}
C_q(b,x,\mu_L^2)&=&\delta(1-x) +
\frac{\alpha_sC_F}{2\pi}\left[(1-x)-\delta(x-1)\left(\frac{\pi^2}{12}\right)\right]
\ ,
\end{eqnarray}
and $C_{\bar q}=C_q$. Note that we have set the quark charge $e_q$
to 1.

\subsection{Resummation and comparison with conventional approach}

The final result for $W(b,Q^2)$ is a combination of the factors we
have calculated in the previous subsections, and as have been
advertised in the introduction \cite{Man03,IdiJi05},
\begin{equation}
W(b, Q^2,x_1,x_2)=\sum_{q\bar q}C^2(Q^2,\alpha_s(Q^2))e^{-{\cal
S}(Q^2,~\mu_L^2)} \left(C_q\otimes
q\right)(x_1,b,Q^2,\mu_L^2)\left(C_{\bar q}\otimes \bar
q\right)(x_2,b,Q^2,\mu_L^2) \ ,
\end{equation}
where the exponential suppression factor can be calculated from
the anomalous dimension calculated in Sec. IIA,
\begin{equation}
{\cal S}=\int_{\mu_L}^Q\frac{d\mu}{\mu}2\gamma_1(Q^2,\mu^2)
=\int_{\mu_L^2}^{Q^2}\frac{d\mu^2}{\mu^2}\left[A\ln\frac{Q^2}{\mu^2}+B\right]\
.
\end{equation}
Here $A$ and $B$ are the single logarithmic and constant terms in
the anomalous dimension $\gamma_1$, respectively. The separation
of the anomalous dimension into a sum of these two terms holds to
any order in perturbation theory. The Sudakov form factor comes
from the running of the matching coefficient $C(Q^2,\mu^2)$ from
$\mu^2=Q^2$ down to the scale of $\mu_L^2$ which can be taken as
$C_1^2/b^2$.

The above is the final resummation result in the effective theory.
$A$ and $B$ coefficient functions can be expanded as a series of
$\alpha_s$:
$A=\sum_{i=1}A^{(i)}\left(\frac{\alpha_s}{\pi}\right)^i$ and
$B=\sum_{i=1}B^{(i)}\left(\frac{\alpha_s}{\pi}\right)^i$. From the
result in Sec. IIA, we have the first two terms of these
expansions,
\begin{eqnarray}
A^{(1)}&=&C_F \ , \nonumber\\
B^{(1)}&=&-\frac{3}{2}C_F \ , \nonumber\\
A^{(2)}&=&\left[\left(\frac{67}{36}-\frac{\pi^2}{12}\right)C_A-
\frac{5}{18}N_f\right]C_F\ , \nonumber\\
B^{(2)}&=&\left(\frac{13}{4}\zeta(3)-\frac{961}{16\times
27}-\frac{11}{48}\pi^2\right)
C_AC_F+\left(\frac{\pi^2}{24}+\frac{65}{8\times
27}\right)N_fC_F\nonumber\\
&&+\left(\frac{\pi^2}{4}- \frac{3}{16}-3\zeta(3)\right)C_F^2   \ .
\end{eqnarray}
In the present formulation, the summation of leading logarithms
(LL) involves $A^{(1)}$; that of next-to-leading (NLL) logarithms
involves $A^{(2)}$ and $B^{(1)}$; and that of
next-to-next-to-leading logarithms (NNLL) involves $A^{(3)}$,
$B^{(2)}$, one-loop $C_q$, and part of the two-loop $C_q$ [we
write $C_q = \sum_n C_q^{(n)}$].

To compare with the CSS approach we follow the procedure outlined
in \cite{CatDefGra01} by absorbing the factor
$C(Q^2,\alpha_s(Q^2))$ into $B$ and $C$ functions, for example, up
to order $\alpha_s$,
\begin{eqnarray}
C_{q/\rm CSS}^{(1)}(x)&=&C_q^{(1)}(x)+\delta(1-x)\frac{1}{2}\left(C(Q^2,\alpha_s(Q^2))\right)^2\ ,\nonumber\\
B_{\rm CSS}^{(2)}&=&B^{(2)}+\beta_0 \frac{\alpha_s}{4\pi}
\left(C(Q^2,\alpha_s(Q^2))\right)^2 \
\end{eqnarray}
At two-loop level and beyond, one has to shuffle the $\ln
Q^2$-dependent part of $C_q$ into $B_{\rm CSS}$ as well. In this
way, we get the CSS resummation formula as,
\begin{equation}
W(Q^2,b^2)=e^{-{\cal S}_{\rm sud}} \left(C_{q/\rm CSS}\otimes
q\right)(x_1,b,\mu_L^2)\left(C_{\bar q/\rm CSS}\otimes \bar
q\right)(x_2,b,\mu_L^2) \ ,
\end{equation}
with
\begin{equation}
{\cal S}_{\rm sud}
=\int_{\mu_L^2}^{Q^2}\frac{d\mu^2}{\mu^2}\left[A_{\rm
CSS}\ln\frac{Q^2}{\mu^2}+B_{\rm CSS}\right]\ .
\end{equation}
$A_{\rm CSS}$ will be the same as our $A$ functions, as does
$B^{(1)}$. For $C_{q/\rm CSS}$, we have
\begin{eqnarray}
C_{q/\rm
CSS}^{(1)}(b,x,\mu_L^2)&=&\frac{\alpha_sC_F}{2\pi}\left[(1-x)-\delta(x-1)\left(4-\frac{\pi^2}{2}\right)\right]\
.
\end{eqnarray}
Comparing with the results in \cite{ColSopSte85}, we find that we
can reproduce the $A^{(1)}_{\rm CSS}$, $A^{(2)}_{\rm CSS}$,
$B^{(1)}_{\rm CSS}$, $C^{(1)}_{\rm CSS}$, which are all the
coefficients and functions needed for resummation at NLL order.
$B^{(2)}_{\rm CSS}$ will be needed to resum NNLL. However, to
fully achieve NNLL resummation we need to calculate the matching
coefficient at the lower scale up to order $\alpha_s^2$ in SCET
\cite{DefGra01}. We leave this to a future publication.

Following the above, the resummation for SIDIS can be performed
similarly. As we stated earlier, the anomalous dimension will be
the same. The only difference is the process-dependent matching
coefficients at $Q^2$ and $\mu_L^2$ in the resummation formula.
For DIS, one has the one-loop result,
\begin{equation}
C_{\rm
DIS}^{(1)}(Q^2,\alpha_s(Q^2))=-\frac{\alpha_s}{2\pi}\left[4-\frac{\pi^2}{12}\right
] \ ,
\end{equation}
which leads to the result for the $C_{q/\rm CSS}^{(1)}$ in DIS,
\begin{eqnarray}
C_{q/\rm CSS}^{(1)}&=&
\frac{\alpha_sC_F}{2\pi}\left[(1-x)-4\delta(x-1)\right]\ .
\end{eqnarray}
These results agree with those from the conventional resummation
approach \cite{{Nadolsky:1999kb}}.

\section{Standard-model Higgs production}

Transverse-momentum dependence of the Standard Model Higgs
production can also be studied through resummation of large double
logarithms. Higgs production, for a large range of Higgs mass, can
be described by an effective action with a pointlike coupling
between the Higgs particle and gluon fields. The effective
coupling is of course scale dependent, balancing the
renormalization dependence of the composite operator. In general,
this effective lagrangian can be written as \cite{vanHiggs},
\begin{equation}
{\cal L}_{Hgg}=C_{EW}(M_t)C_T(M_t,\mu)H F^{a}_{\mu\nu}F^{a\mu\nu}
\ ,
\end{equation}
where $H$ is the scalar Higgs field and $F^a_{\mu\nu}$ is the
gluon field strength tensor. $C_{EW}$ represents the electroweak
coupling coefficient from the heavy-top-quark loop calculation,
while $C_T$ comes from the strong interaction radiative
corrections. The coefficient $C_T$ will depend, in general, on the
top quark mass and the renormalization scale $\mu$. To our
interest, we quote at two-loop
\cite{HiggsOneloop,{Chetyrkin:1997iv}},
\begin{eqnarray}
C_T(M_t,\mu)&=&\frac{\alpha_s(\mu)}{4\pi}\left\{1+\frac{\alpha_s(\mu)}{4\pi}\left(5C_A-3C_F\right)\right.\nonumber\\
&&\left.+\frac{\alpha_s^2(\mu)}{(4\pi)^2}\left[\ln\frac{\mu^2}{M_t^2}\left(7C_A^2-11C_AC_F+8N_fT_FC_F\right)+\cdots\right]
\right\}\ . \label{ct}
\end{eqnarray}
Here we have omitted the constant terms of order $\alpha_s^2$
because they do not contribute to the renormalization group
running at two-loop order.

In SCET, the Higgs production cross section can be calculated from
its coupling with the collinear gluon fields,
\begin{equation}
{\cal L}_{H-SCET}=C(M_H,\mu)H {\cal F}^{a}_{n\mu\nu}{\cal
F}^{a\bar n\mu\nu} \ ,
\end{equation}
where ${\cal F}^{a}_{n,\bar n \mu\nu}$ represent the $n$ and $\bar
n$ collinear gluon field strength tensors in SCET \cite{SCET}.
$C(M_H,\mu)$ is the matching coefficient which contains the
coupling of Higgs boson to gluons in full QCD,
\begin{equation}
C(M_H,\mu)=C_{EW}(M_t)C_T(M_t,\mu)C_G(M_H,\mu) \ .
\end{equation}
The last factor comes from matching between operator $H
F^{a}_{\mu\nu}F^{a\mu\nu}$ in full QCD and $H{\cal
F}^{a}_{n\mu\nu}{\cal F}^{a\bar n\mu\nu}$ in SCET. Because
$C_{EW}$ has no QCD effects, we will not discuss it further in
this paper. $C_T$ and $C_G$ contain the QCD evolution effects, and
thus the anomalous dimension of the SCET operator $H{\cal
F}^{a}_{n\mu\nu}{\cal F}^{a\bar n\mu\nu}$ is the sum of the two:
\begin{equation}
\gamma_1=\gamma_T+\gamma_G \ ,
\end{equation}
where $\gamma_T$ and $\gamma_G$ are defined as
\begin{eqnarray}
\gamma_T&=&\frac{d \ln C_T(\mu)}{ d \ln\mu}\ ,
\nonumber\\
\gamma_G&=&\frac{ d \ln C_G(\mu)}{ d \ln\mu}\ .
\end{eqnarray}
From Eq.~(\ref{ct}), it is easy to show that
\begin{eqnarray}
\gamma_T &=&
-2\beta_0\frac{\alpha_s}{4\pi}+\frac{\alpha_s^2}{(4\pi)^2}\left[-2\beta_1-2\beta_0(5C_A-3C_F)
 \right. \nonumber \\
&&\left. +2\left(7C_A^2-11C_AC_F+8N_fT_FC_F\right)\right] \ ,
\end{eqnarray}
up to two-loop order. To calculate $\gamma_G$, we follow the
calculation for the Drell-Yan process in the previous section.
Using the gluon form factor in \cite{vanHiggs,Har00}, we get the
relevant anomalous dimension,
\begin{eqnarray}
\gamma_G&=&\frac{\alpha_s}{\pi}C_A\ln\frac{M_H^2}{\mu^2}\nonumber\\
      &&+\frac{\alpha_s^2}{(4\pi)^2}\left[\frac{1}{2}C_A
K\ln\frac{M_H^2}{\mu^2}-\gamma_{gg}^{(1)}-2f_{g2}^{(1)}+4\beta_1\right]
\ ,
\end{eqnarray}
where $f_{g2}^{(1)}$ can be obtained from the above $f_{q2}^{(1)}$
by color-factor exchange ($C_A\leftrightarrow C_F$)
\cite{vanHiggs}. The anomalous dimension $\gamma_{gg}^{(1)}$
\begin{equation}
\gamma_{gg}^{(1)}=C_A^2\left(\frac{64}{3}+24\zeta(3)\right)-\frac{32}{3}N_fT_FC_A-
8N_fT_FC_F \ ,
\end{equation}
is proportional to $\delta(1-x)$ in the gluon splitting function.
$\beta_1$ is the two-loop beta function defined before.

Including the matching at lower scale $\mu_L$, the result for
Higgs production $W(b,M_H^2,x_1,x_2)$ at low transverse momentum
can be written as,
\begin{eqnarray}
W(M_H^2,b)=C^2_H(M_H^2,\alpha_s(M_H^2))e^{-{\cal
S}}\left(C_g\otimes g\right)(x_1,b,\mu_L^2)\left(C_{ g}\otimes
g\right)(x_2,b,\mu_L^2)\ ,
\end{eqnarray}
where $C_{EW}$ and the leading factor in $C_T$ have been absorbed
in the Born cross section, and the remainder $C_H$ is
\begin{equation}
C_H(M_H^2,\alpha_s(M_H^2))=1+\frac{\alpha_s(M_H^2)}{4\pi}\left[5C_A-3C_F+C_A\left(\zeta(2)+\pi^2\right)\right]
\ ,
\end{equation}
up to one-loop order. The Sudakov suppression form factor has the
same form as that for DY, and the expansions of the $A$ and $B$
are,
\begin{eqnarray}
A_H^{(1)}&=&C_A \nonumber\\
B_H^{(1)}&=&-\frac{1}{2}\beta_0\nonumber\\
A_H^{(2)}&=& C_A K\nonumber\\
B_H^{(2)}&=&-\frac{1}{16}\left[\gamma_{gg}^{(1)}+2f_{g2}^{(1)}+2\beta_0\left(5C_A-3C_F\right)-2\beta_1
\nonumber\right. \\
&&\left.- 2\left(7C_A^2-11C_AC_F+8N_fT_FC_F\right)\right] \
\end{eqnarray}
With these results we can reproduce the conventional resummation
for Higgs-boson production at NLL order \cite{DefGra01}.

The coefficient $C_g$ to one-loop order is calculated from the
matching at the scale of $\mu_L$, with real and virtual
contributions. It is needed for resummation at NLL. The result is
\begin{eqnarray}
C_g(b,x,\mu_L^2)&=& \delta(1-x) +
\frac{\alpha_sC_A}{2\pi}\delta(x-1)\left(-\frac{\pi^2}{12}\right)
\ ,
\end{eqnarray}
where we have chosen $\mu_L=C_1/b$ with $C_1=2e^{-\gamma_E}$ to
eliminate the large logarithms. The corresponding coefficient for
CSS resummation is
\begin{equation}
C_{g/\rm CSS}(b, x, \mu_L^2) = \delta(1-x)
 + \frac{\alpha_sC_A}{4\pi}\delta(1-x)\left[ (5+ \pi^2)C_A - 3
 C_F\right]
\end{equation}
which is consistent with \cite{DefGra01}.

\section{Conclusion}

In this paper, we demonstrated how to perform resummation of large
double logarithms in hard processes involving low transverse
momentum in the framework of effective field theory. The
newly-discovered theory, SCET, is naturally suited for various
hard processes with two or more well-separated momentum scales .
As an example, we studied the Drell-Yan resummation in detail, and
outlined how to extend the procedure to other processes as well.
We calculated the relevant anomalous dimensions of the effective
current up to two-loop order, and reproduced the conventional
resummation results to NLL accuracy. This method can be certainly
extended to higher order as well, for example, NNLL. However, this
requires calculation of the matching coefficient at the
intermediate scale within SCET up to two-loop order. We leave
these extensions for a future publication.

\section*{Acknowledgments}
A. I. and X. J. are supported by the U.S. Department of Energy via
grant DE-FG02-93ER-40762. F.Y. is grateful to RIKEN, Brookhaven
National Laboratory and the U.S. Department of Energy (contract
number DE-AC02-98CH10886) for providing the facilities essential
for the completion of his work.

\end{document}